\documentclass[structabstract,usenatbib,useAMS]{aa}  
\usepackage{graphicx}
\usepackage{txfonts}
\usepackage{natbib}
\usepackage{amssymb}
\usepackage{subfigure}
\usepackage{multirow}

\usepackage{txfonts}
\begin{document}
   \title{Revisiting the Impact of Atmospheric Dispersion and Differential Refraction on Widefield Multiobject Spectroscopic Observations}
   
   \titlerunning{Atmospheric dispersion in widefield MOS observations}

   \subtitle{From VLT/VIMOS to Next Generation Instruments}

   \author{R. S\'anchez-Janssen
          \inst{1,2}
          \and
          S. Mieske
          \inst{2}
          \and
          F. Selman
          \inst{2}
          \and
          P. Bristow
	  \inst{3}
          \and
          P. Hammersley
          \inst{3}
          \and
          M. Hilker
          \inst{3}
          \and
          M. Rejkuba
          \inst{3}
          \and
          B. Wolff
	\inst{3}
          }

   \institute{NRC Herzberg Institute of Astrophysics, 5071 West Saanich Road, Victoria V9E2E7, Canada\\
              \email{ruben.sanchez-janssen@nrc-cnrc.gc.ca}
              \and European Southern Observatory, Alonso de C\'ordova 3107, Santiago, Chile
              \and European Southern Observatory, Karl-Schwarzschild-Strasse 2, D-85748 Garching bei M\"unchen, Germany
                   }
                   
   \date{Received ...; accepted ...}

 
  \abstract
   {Atmospheric dispersion and field differential refraction impose severe constraints on widefield, multiobject spectroscopic (MOS) observations, where the two joint effects cannot be continuously corrected. Flux reduction and spectral distortions must then be minimised by a careful planning of the observations -- which is especially true for instruments that use slits instead of fibres.
   This is the case of VIMOS at the VLT, where MOS observations have been restricted, since the start of operations, to a narrow two-hour range from the meridian to minimise slit losses -- the so-called two-hour angle rule. } 
   {We revisit in detail the impact of atmospheric effects on the quality of VIMOS-MOS spectra with the aim of enhancing the instrument's overall efficiency, and improving the scheduling of observations.}
   {We model slit losses across the entire VIMOS field-of-view as a function of target declination. We explore two different slit orientations at the meridian: along the parallactic angle (North-South), and perpendicular to it (East-West).}
   {We show that, for fields culminating at zenith distances larger than 20 deg, slit losses are minimised with slits oriented along the parallactic angle at the meridian.
The two-hour angle rule holds for these observations using North-South orientations.
Conversely, for fields with zenith angles smaller than 20 deg at culmination, losses are minimised with slits oriented perpendicular to the parallactic angle at the meridian. MOS observations can be effectively extended to plus/minus three hours from the meridian in these cases. In general, night-long observations of a single field will benefit from using the East-West orientation.
All-sky or service mode observations, however, require a more elaborate planning that depends on the target declination, and the hour angle of the observations.
}
   {We establish general rules for the alignment of slits in MOS observations that will increase target observability, enhance the efficiency of operations, and speed up the completion of programmes -- a particularly relevant aspect for the forthcoming spectroscopic public surveys with VIMOS.
   Additionally, we briefly address the (non-negligible) impact of field differential refraction on future widefield MOS surveys.}


   \maketitle
%

\section{Introduction}

\begin{table}[!h]
\caption{Characteristics of operational and planned/proposed multiobject optical spectrographs with FOVs larger than 100 arcmin$^2$.}
\begin{tabular}{llccccc} 
\hline\hline
Telescope & Instrument  & o/p\,\tablefootmark{a}  & Aperture & s/f\,\tablefootmark{b} &  FOV & N$_{max}$\\
 &   &  & [m] &  &  [deg$^2$] &  \\
\hline
SDSS	&    	SDSS     		&   o	  &	2.5	  &     f 	&	7.07		&	1000\\
WIYN	&   	HydraW		&   o   &	3.5	  &     f	&	0.78		&	100\\
Blanco	&   	HydraB		&   o   &	3.9	  &     f	&	0.35		&	138\\
AAT	    	&	2dF		 	&   o   &	3.9	  &     f	&	3.14		&	392\\
LAMOST	&   	LAMOST		&   o   &	4.0	  &     f	&	19.6		&	4000\\
WHT	    	&	WYFFOS		&   o   &	4.2	  &     f	&	0.35		&	150\\
Magellan   &	IMACS		&   o   &	6.5	  &     s	&	0.2		&	600\\
MMT	    	&	Hectospec		&   o   &	6.5	  &     f	&	0.78		&	300\\
VLT	    	&	VIMOS		&   o   &	8.2	  &     s	&	0.08		&	600\\
VLT	    	&	FLAMES		&   o   &	8.2	  &     f	&	0.14		&	132\\
Mayall	&   	DESI			&   p   &	3.9	  &     f	&	7.07		&	5000\\
VISTA	&   	4MOST		&   p   &	4.1	  &     f	&	7.07		&	3000\\
WHT	    	&	WEAVE		&   p   &	4.2	  &     f	&	3.14		&	1000\\
VLT	    	&	MOONS		&   p   &	8.2	  &     f	&	0.14		&	500\\
Subaru	&   	PFS			&   p   &	8.3	  &     f	&	1.33		&	2400\\  
ngCFHT	&   	ngCFHT		&   p   &	10.0	  &     f	&	1.5		&	4000\\
GMT	    	&	GMACS		&   p   &	21.9	  &     s	&	0.05	&	400\\
\hline
\end{tabular}
\label{table:facilities}
\tablefootmark{a}{(o)perational/(p)lanned/(p)roposed}\\
\tablefootmark{b}{(s)lits/(f)ibres}\\
\end{table}

The effects of atmospheric dispersion on spectrophotometric observations were first tackled in a seminal paper by \citet{Filippenko1982}, and have since been addressed by many other authors \citep{Cohen1988,Donnelly1989,Cuby1994,Szokoly2005}. 
Two different components contribute to slit losses: a chromatic dispersion caused by the wavelength variation of the index of refraction of air; and an achromatic differential refraction due to airmass variations across the field-of-view (FOV).
As pointed out by \citet{Cuby1998}, the chromatic effect is almost constant for a given field, and can thus be counterbalanced with an atmospheric dispersion compensator (ADC). 
On the other hand, field differential refraction cannot be continuously corrected, and this is especially problematic for optical instruments with large FOVs. In the case of multiobject spectroscopic (MOS) observations the two joint effects cannot be compensated, so that aperture losses must be minimised by a careful planning of the observations \citep[e.g.,][]{Cuby1998,Szokoly2005}. This may range from frequent reconfiguration of the fibres, to imposing limited observability windows.
The latter is actually the only alternative for instruments that use slits instead of fibres, even more so because field rotation prevents the alignment of all slits along the parallactic angle.

\begin{figure}
\centering
\includegraphics[width=.5\textwidth]{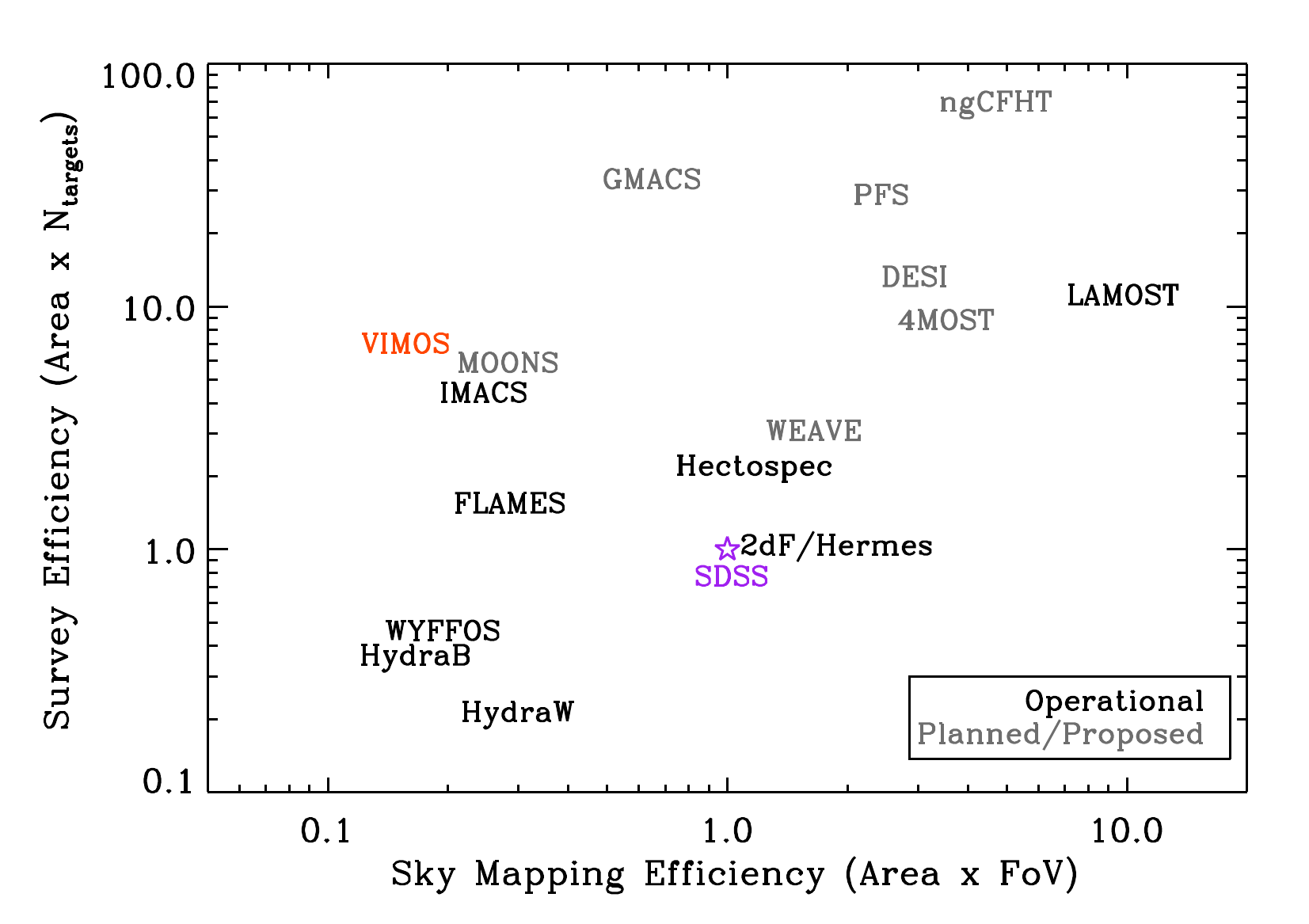}
\caption{Survey efficiency (telescope area $\times$ maximum number of allocated targets) and sky mapping efficiency (telescope area $\times$ instrument field-of-view, or \'etendue) of all operational (black), and planned/proposed (grey) multiobject optical spectrographs with FOVs larger than 100 arcmin$^2$. All figures are normalised to those of the SDSS. Note the extraordinary survey efficiency of VIMOS, only comparable to that of next generation instruments.}
\label{fig:vimos}
\end{figure}

VIMOS\,\footnote{http://www.eso.org/sci/facilities/paranal/instruments/vimos.html} \citep{LeFevre2003} is a widefield (4x7x8 arcmin$^{2}$) instrument with imaging, integral field, and MOS capabilities mounted at the Nasmyth B focus of VLT UT3. 
The instrument operates in the optical wavelength range (360-1000 nm), and is equipped with six sets of grisms, six sets of broad-band filters, plus three additional filter sets specifically designed to be used in combination with the grisms to block the second order spectra.
The unique combination of instrument FOV, large collecting power, and very high multiplexing (up to $\sim$\,600 targets in the low resolution modes; see Table\,\ref{table:facilities} and Fig.\,\ref{fig:vimos}) has made VIMOS a particularly efficient instrument for large spectroscopic surveys of cosmological fields \citep[e.g.,][]{LeFevre2005,Lilly2007,Popesso2009,Guzzo2013}.

VIMOS-MOS observations are carried out using multislit masks, which provide very accurate sky subtraction and high instrumental throughput.  However, the lack of ADCs makes atmospheric dispersion and field differential refraction factors that need to be taken into account.
The atmospheric effects in instruments like VIMOS were studied by \citet{Cuby1998}. 
They show that i) atmospheric dispersion dominates at shorter wavelengths, while differential refraction is not negligible at the red end of the visible spectrum; ii) in the former case, image drifts occur along the meridian; iii) differential refraction can be almost neglected for zenith distances smaller than 25 deg for exposure times up to two hours from the meridian (HA = 0\,h).
In view of these results they recommended that, in order to minimise slit losses, the slits be positioned along the dispersion direction at mid-exposure, and observations be limited to a narrow two-hour range from meridian crossing -- the so-called two-hour angle rule. This guarantees that losses remain below 20 per cent for zenith angles $<50$ deg at culmination.
These rather limiting guidelines had always been in place for all MOS observations since the start of operations in 2003, and in practice translate into mandatory airmass constraint limits for the VIMOS-MOS observing blocks (OBs).

During the last few years the instrument performance has been significantly enhanced (see \citealt{Hammersley2010,Hammersley2013}) by changing the detectors to red-sensitive, low-fringing CCDs; replacing the HR-blue grism set with higher throughput VPH grisms; introducing an active flexure compensation system; redesigning the focusing mechanism and mask cabinet; and introducing a new pre-image-less MOS mode \citep{Bristow2012}. All these improvements have made VIMOS a much more stable instrument, and have extended its lifetime in order to prepare it for the start of the spectroscopic public surveys for which ESO has recently issued a call.
Further work to improve the operational efficiency of the instrument includes the present study, which has as main goal to revisit the need for restricted observability of targets only within plus/minus two hours from the meridian in the MOS mode.
Increasing the observability of targets in the MOS mode provides more flexibility to operations, because the number of masks that can be loaded in the instrument before the beginning of each night is limited.
As we previously noted, VIMOS is very often used for deep observations of cosmological fields, where very long integrations are taken for the same field.
By increasing the target visibility (relaxing the two-hour angle rule), the programmes can be completed faster.

In this paper we revisit in detail the impact of atmospheric dispersion on the quality -- in terms of slit losses and spectrophotometric distortions -- of VIMOS observations. 
Our aim is to establish general rules for the alignment of slits in MOS observations that enhance the efficiency of operations, not only for VIMOS, but for other operational and future multislit spectrographs as well.
We note that the parameter space of this problem is large. Irrespective of image quality and atmospheric conditions, slit losses depend on slit size, orientation, and position within the FOV, observed wavelength range, target declination, total exposure time, and hour angle (HA) of the observations. 
We build upon the previous work by \citet{Cuby1998}, but include specific calculations for all the grisms currently available in VIMOS, and present a more detailed analysis of the optimal slit orientation as a function of target declination.

This paper is organised as follows. 
In Section\,\ref{sect:model} we describe the model with which we compute slit losses in VIMOS.
Section\,\ref{sect:slitlosses} presents the main results of our analysis, while in Section\,\ref{sect:fdr} we briefly address the specific contribution of field differential refraction, and its importance for widefield MOS observations.
Finally, in Section\,\ref{sect:conclusions} we discuss our main results, and summarise our findings.

%
%

\section{A model to address slit losses in VIMOS-MOS}\label{sect:model}

In order to investigate potential operational improvements that could enhance the efficiency of VIMOS, we have simulated the effects of slit losses under different circumstances. 
Our fiducial model (see Table\,\ref{table:model}) assumes a flat input spectrum and nine slits evenly distributed across the entire VIMOS FOV, from the centre to the corners, and with relative separations of seven arcmin.
All slits have a $l=10$ arcsec length and $w=1$ arcsec width, which is typical for the majority of VIMOS-MOS observations.
We assign two different orientations for the slits at meridian crossing, namely North-South (PA = 0 deg), and East-West (PA = 90 deg).\,\footnote{Please note that here we follow the usual on-sky convention for orientations. This is therefore different from the rotator offset angle described in the VIMOS manual: the default offset angle of 90 deg corresponds to a N-S on-sky orientation, while a nonstandard rotator offset angle of 0 deg corresponds to E-W on sky.} 
\citet{Szokoly2005} suggests that, under specific circumstances (e.g., night-long observations of a given field, or observations that extend further into the red end of the visible spectrum), the best option is the latter -- to orient the slits \emph{perpendicular} to the parallactic angle at meridian --, and we therefore explore both possibilities.
Alignment and guiding -- in the centre of the FOV -- are assumed to be done at either 450 nm (for the blue grisms only; see Table\,\ref{table:vimos_setups}) or 700 nm (for the rest). The seeing PSF is considered to be wavelength- and airmass-independent, with a Gaussian FWHM = 1 arcsec.
In reality, of course, the seeing will vary as a function of both wavelength ($\lambda$) and airmass ($X$), FWHM $\propto \lambda^{-1/5}\,X^{3/5}$ \citep{Kolmogorov1941}. 
But given that during service-mode operations at ESO seeing constraints have to be satisfied at any given airmass and instrument setup, the assumption that the seeing PSF is independent of wavelength and airmass in the model is justified, because the observations are always scheduled such that this is in fact realised.
This setup results in a 24 per cent fiducial flux loss due to finite seeing and slit width, and under the assumption that the objects are perfectly centred within the slits at the beginning of the observations.
We adopt the average night-time pressure (743 mbar) and temperature (12 $^{\circ}$C) at the Paranal Observatory within the last five years (J. Navarrete, private communication) in our computation of atmospheric dispersion, for which we follow \citet{Filippenko1982}.
Finally, for each of the six VIMOS grisms (and filter combinations; Table\,\ref{table:vimos_setups}) we assume observations with 3600 s exposure time, within four hours from meridian crossing, and for targets  in the $-75 \leq \delta \leq +25$ deg declination range. All these parameters are typical of VIMOS-MOS service-mode observations.

 \begin{table}[!h]
\centering
\caption{Parameters of the slit losses model.}
\begin{tabular}{ll} \hline\hline
\multicolumn{1}{c}{Parameter} & \multicolumn{1}{c}{Value}\\ \hline
\multicolumn{2}{c}{\emph{Site Properties}} \\
Paranal Latitude & $\varphi = -24.5$ deg \\
Atmospheric Pressure & P = 743 mbar\\
Temperature & T = 12 $^{\circ}$C\\
Seeing & FWHM = 1 arcsec\\
\hline
\multicolumn{2}{c}{\emph{Observational Setup}} \\
Slit Width & $w = 1$ arcsec\\
Slit Length & $l = 10$ arcsec\\
\multirow{2}{*}{Slit Orientation} & PA = 0 deg (North-South)\\
 &  PA = 90 deg (East-West)\\
 Target Declination & $-75 \leq \delta \leq +25$ deg\\
 Hour Angle & $0 \leq\,\mid$HA$\mid\,\leq 4$ h\\
 Exposure Time & 3600 s\\
\hline
\multicolumn{2}{c}{\emph{Figures of Merit}} \\
Spectral Distortion & $\Delta = 1 - f_{o,min}/f_{o,max}$\\
Relative Flux Loss & $f = 1 -  [\int_{\lambda_{min}}^{\lambda_{max}} \! f_{o}(\lambda)~/ \int_{\lambda_{min}}^{\lambda_{max}} \! f_{i}]$\\
\hline
\multicolumn{1}{c}{}\\
\end{tabular}
\label{table:model}
\end{table}

\begin{table}[!h]
\centering
\caption{VIMOS-MOS setups}             
\begin{tabular}{l l r r r c}        
\hline\hline                 
Grism & Filter & $\lambda_{min}$ & $\lambda_{max}$ & R~~~~ & Dispersion\\    
  & & (nm) & (nm) & (1" slit) & (\AA/pix)\\
\hline                        
HR\_blue & Free  & 370 & 535 & 1150 & 0.7\\
LR\_blue & OS-blue & 370 & 670 & 180 & 5.3\\
HR\_orange &	GG435 & 515 & 760 & 2150 & 0.6\\
MR & GG475 & 480 & 1000 & 580 &	2.5\\
HR\_red & GG475 & 650 & 875 & 2500 & 0.6\\
LR\_red  & OS-red & 550 & 950 & 210 & 7.3\\
\hline                                   
\end{tabular}
\label{table:vimos_setups}      
\end{table}

\subsection{Deviations from the fiducial model}\label{sect:deviations}

\begin{figure}\centering
\includegraphics[width=.5\textwidth]{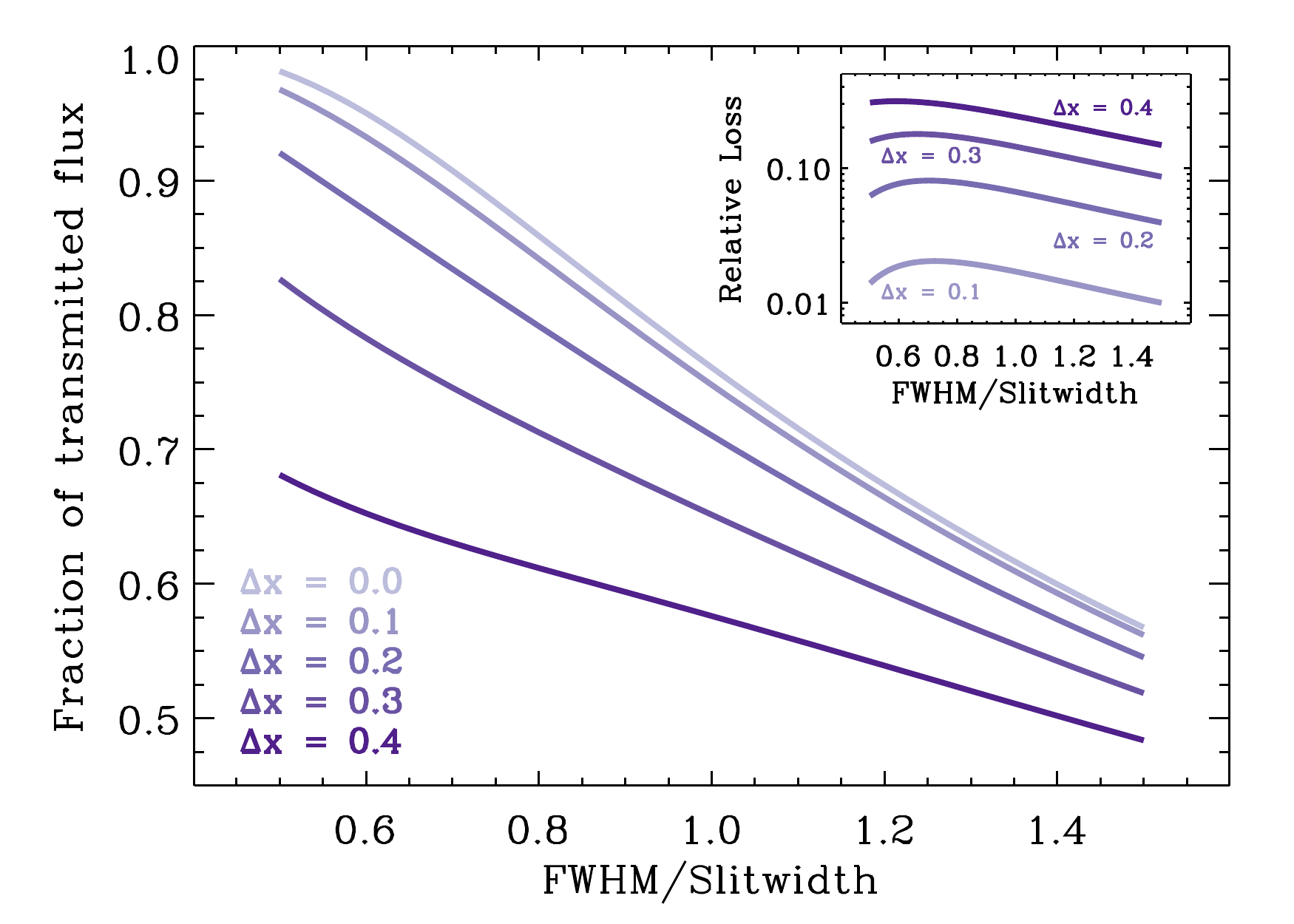}
\caption{Fraction of transmitted flux as a function of FWHM-to-slitwidth ratio, for a Gaussian point source that is displaced perpendicular to the slit by a fractional amount of the slitwidth, $\Delta x = (x-x_{0})/w$. The inset panel shows the flux loss relative to the case where the object is perfectly centred within the slit. A slitlength $l=10\,w$ is assumed.}
\label{fig:slit_flux}
\end{figure}

We previously noted that the parameter space of this problem is very large. As a result, our fiducial model necessarily contains several simplifications, including a fixed FWHM-to-slitwidth ratio, and the assumption that the objects are perfectly centred within the slits. 
Here we estimate, to first order, how the results would change if we modify these conditions.
The first parameter we vary is the FWHM-to-slitwidth ratio.
Ideally, one would optimise the observations by adapting the slitwidth to the seeing conditions. While this ratio can of course not be anticipated in actual observations, the users can significantly restrict it by imposing a given seeing constraint in their service-mode OBs. In any case, we explore a sensible range of FWHM-to-slitwidth ratios, between 0.5 and 1.5.
The second condition we relax is the assumption that the objects are perfectly centred within the slits.
Typically, there is a lower limit of one pixel\,\footnote{0.205 arcsec for VIMOS.} to the precision of mask positioning due to a variety of effects -- including mechanical hysteresis, blind corrections during exposures, etc.
Additional errors due to different airmass/HA between pre-imaging and MOS observations (or even pre-imageless MOS mask creation) will introduce an additional source of flux reduction.
We estimate this flux reduction by assuming the object is displaced perpendicular to the slit -- i.e., in the dispersion direction --, and compute the fractional through-slit flux that is transmitted in each case. 
We note that this calculation ignores all atmospheric dispersion and differential refraction effects. It is therefore equivalent to the flux obtained at the centre of the FOV, for a field observed at zenith.

\begin{figure*}\centering
\includegraphics[width=.9\textwidth]{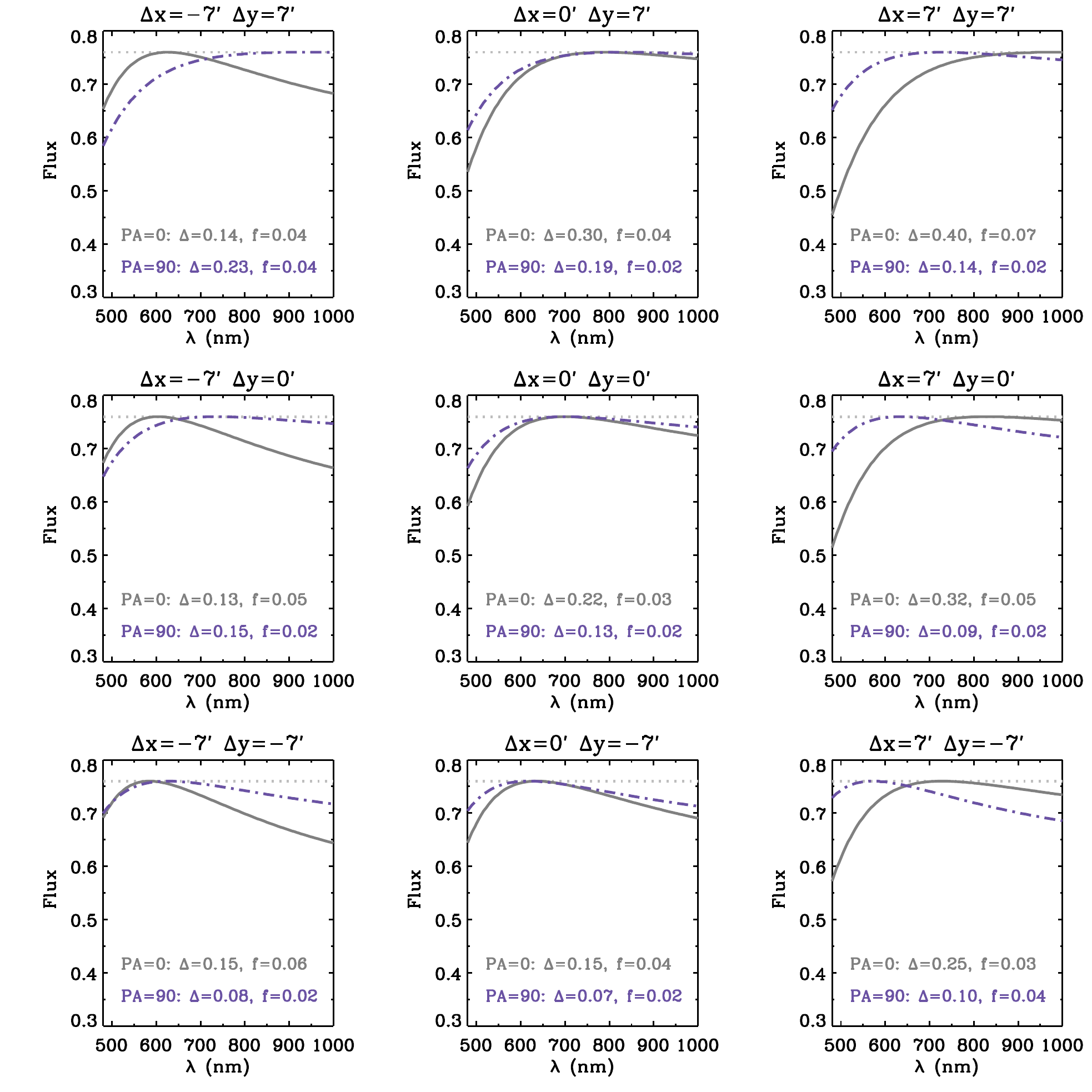}
\caption{Output simulated spectra for nine different slit positions across the VIMOS FOV.
The relative positions (in arcmin) of the slits  with respect to the centre of the FOV are indicated. At the meridian, $\Delta$x and $\Delta$y correspond to $\Delta\alpha$ and $\Delta\delta$, respectively.
These are the result of a one-hour-long integration (-3 $<$ HA (h) $<$ -2) on a $\delta = 0$ deg field using the MR grism. In each panel we show the output spectra for two different slit orientations (solid for N-S,  and dot-dashed for E-W). We also provide the corresponding relative flux loss ($f$) and spectral distortion ($\Delta$). The input spectrum is flat (dotted lines), but its flux is reduced by 24 per cent due to finite seeing and slit width. See text for details.}
\label{fig:slitloss_ex}
\end{figure*}

In Figure\,\ref{fig:slit_flux} we show the fraction of transmitted flux as a function of FWHM-to-slitwidth ratio, for a Gaussian point source that is displaced perpendicular to the slit by a fractional amount of the slitwidth, $\Delta x = (x-x_{0})/w$. 
For VIMOS, and in the particular case of a $w=1$ arcsec slit, the selected displacements correspond to $0 \leq \Delta x \leq 2$ pixels.
The inset panel shows the fractional flux reduction relative to the case where there is no displacement, i.e., $\Delta x = 0$.
Fig.\,\ref{fig:slit_flux} shows the importance of accurate centroiding, as significant losses can otherwise occur. In the case of VIMOS, however, a displacement of up to one pixel only results in a relative flux reduction of $\approx 6$ per cent. 
We recall that, for given conditions, this flux reduction is an absolute minimum, purely a result of geometrical effects. As we will see, slit losses will generally increase (nonlinearly) as a result of atmospheric dispersion and differential refraction during the observations.

%
%

\section{VIMOS-MOS slit losses: dependence on grism, FOV orientation, field declination, and HA}\label{sect:slitlosses}

\begin{figure*}\centering
\includegraphics[width=.98\textwidth]{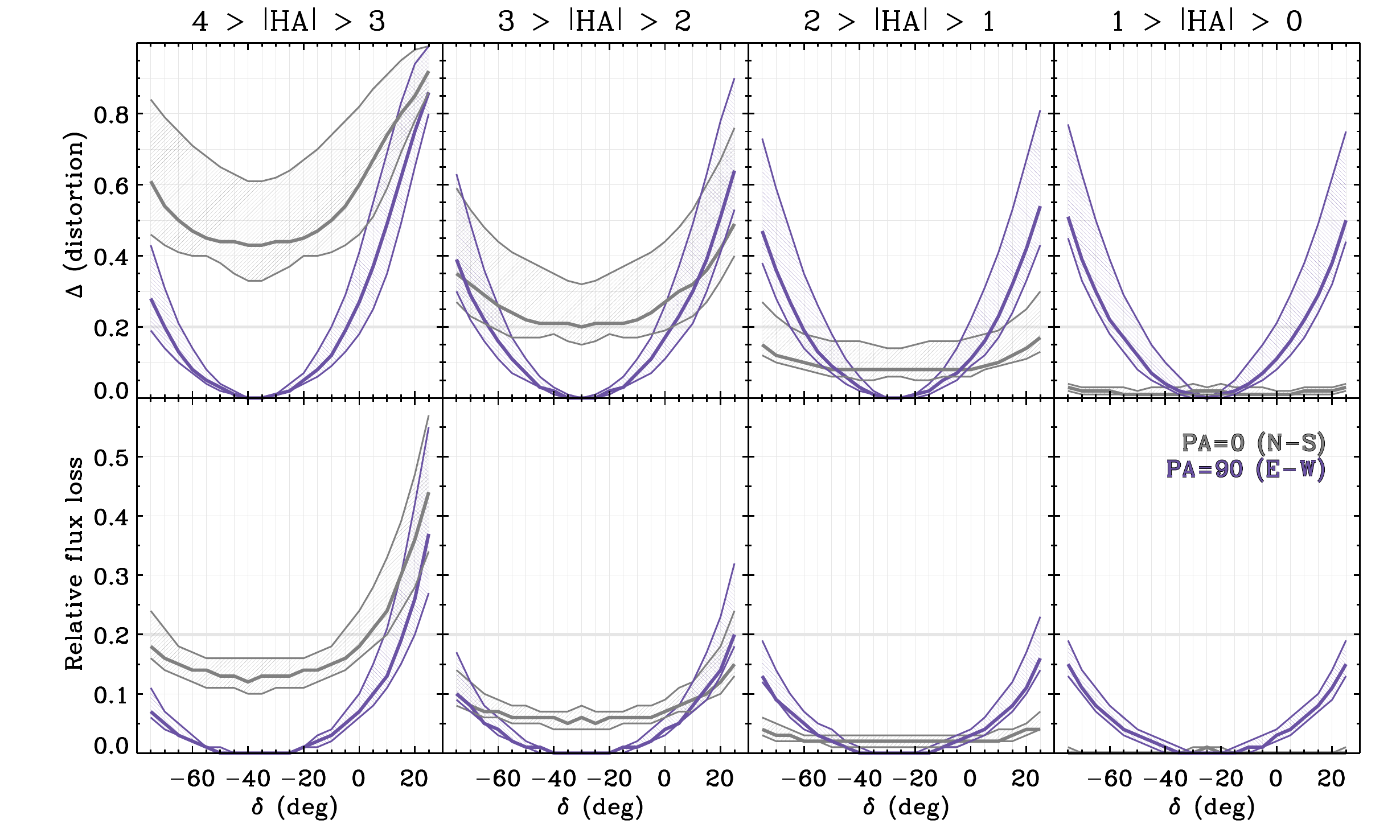}
\caption{Shaded regions show,  for the HR-blue grism, the entire range of spectral distortions (upper row) and flux losses (lower row) for the nine simulated slits as a function of target declination.  The corresponding minimum, median and maximum values at fixed declination are indicated by solid lines. The plots show the effects for two different slit orientations at the meridian (N-S in grey and E-W in purple). Each column corresponds to a one-hour-long integration with target hour angle as indicated on top. The thick horizontal lines at 20 per cent indicate our assumed maximum tolerated loss/distortion level.
}
\label{fig:hrblue}
\end{figure*}

\begin{figure*}\centering
\includegraphics[width=.98\textwidth]{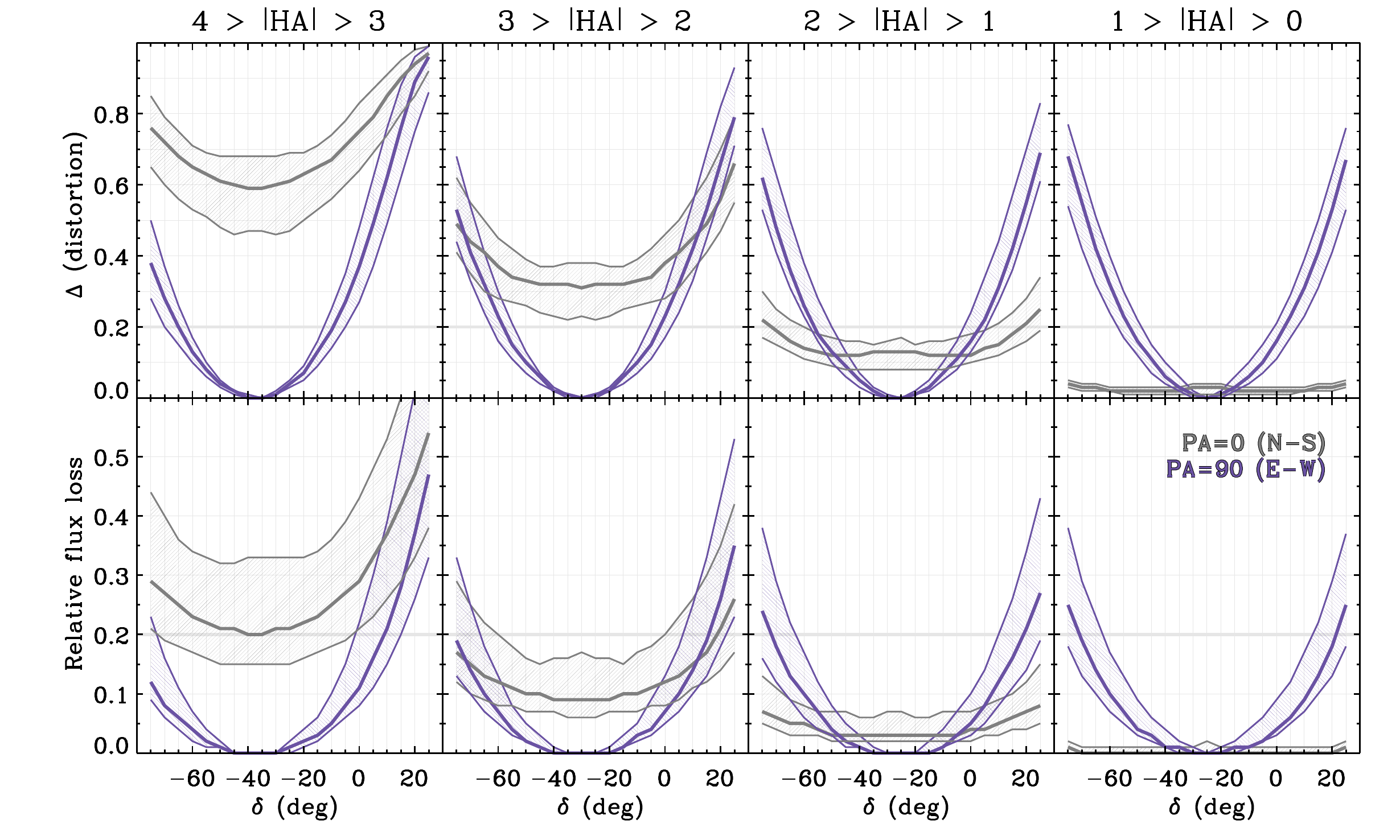}
\caption{Same as Fig.\,\ref{fig:hrblue}, but for the LR\_blue grism.}
\label{fig:lrblue}
\end{figure*}

\begin{figure*}\centering
\includegraphics[width=1.\textwidth]{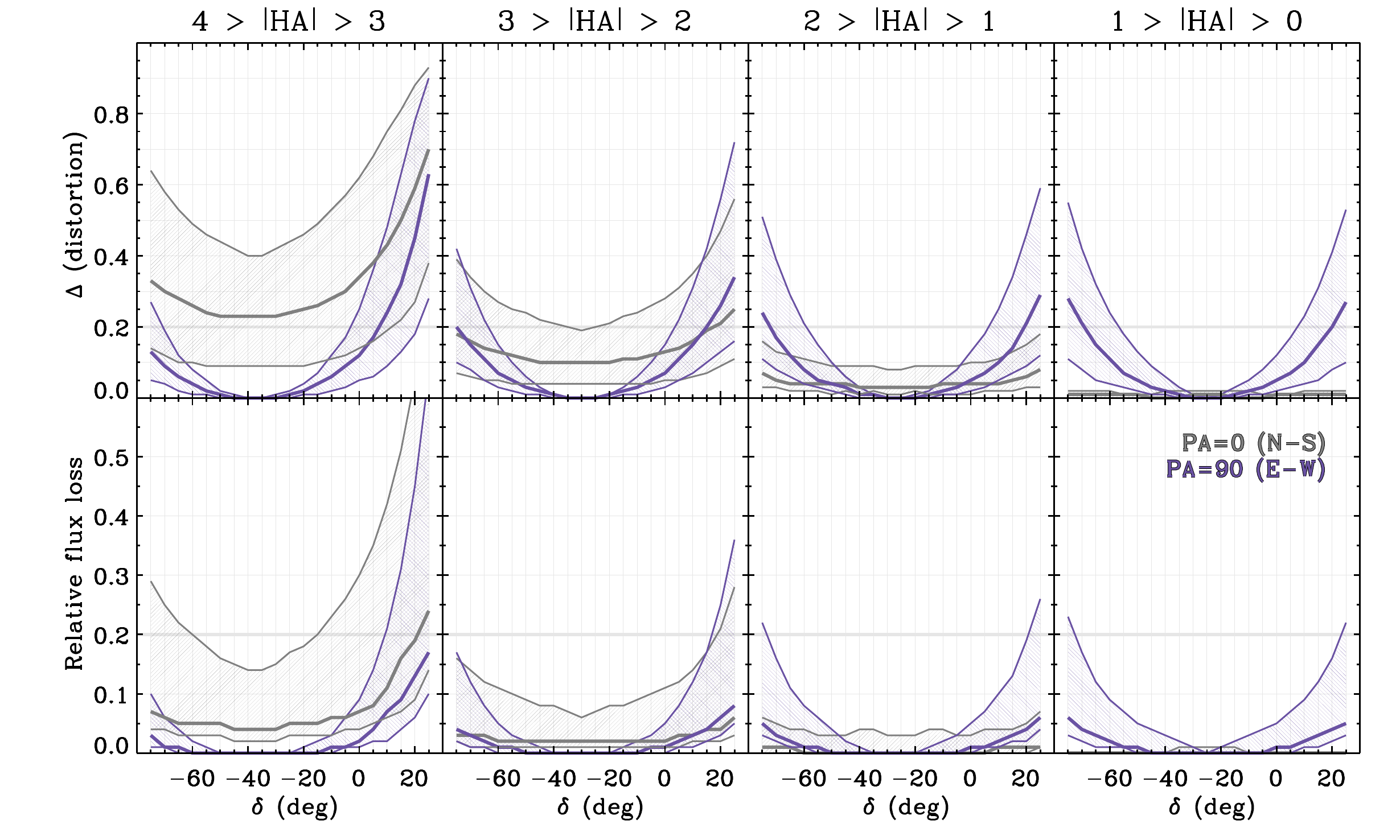}
\caption{Same as Fig.\,\ref{fig:hrblue}, but for the HR\_orange grism.}
\label{fig:hroran}
\end{figure*}

\begin{figure*}\centering
\includegraphics[width=1.\textwidth]{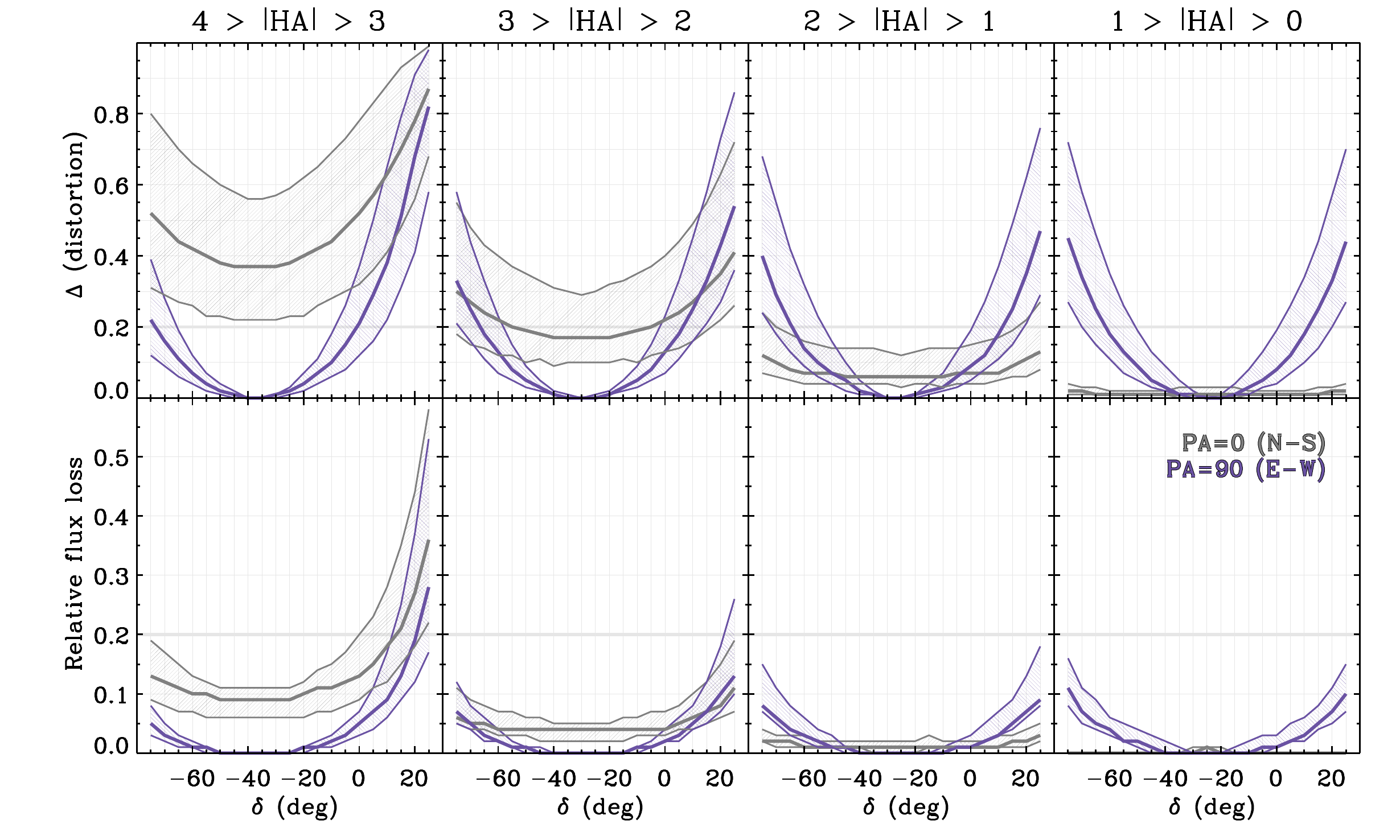}
\caption{Same as Fig.\,\ref{fig:hrblue}, but for the MR grism.}
\label{fig:mr}
\end{figure*}

\begin{figure*}\centering
\includegraphics[width=1.\textwidth]{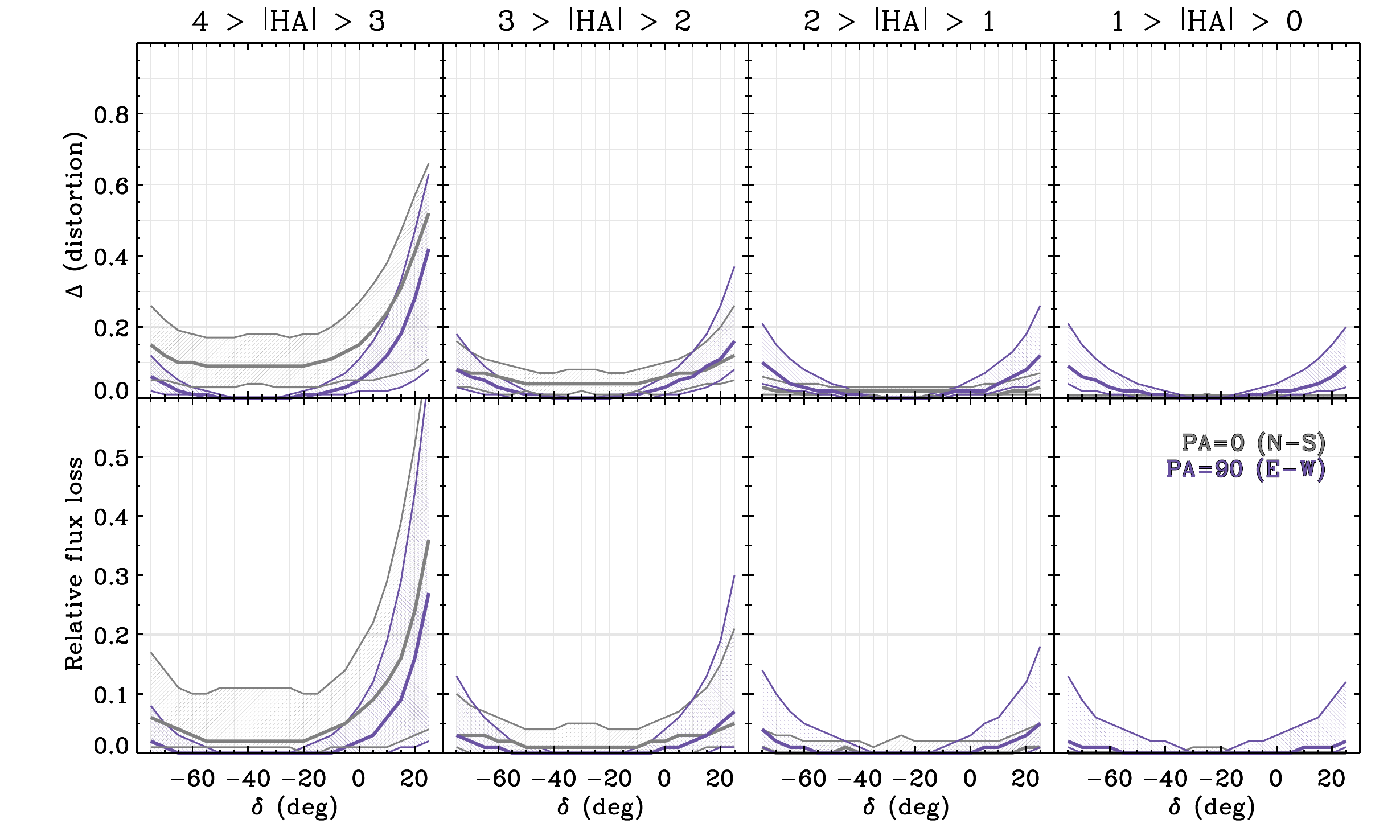}
\caption{Same as Fig.\,\ref{fig:hrblue}, but for the HR\_red grism.}
\label{fig:hrred}
\end{figure*}

\begin{figure*}\centering
\includegraphics[width=1.\textwidth]{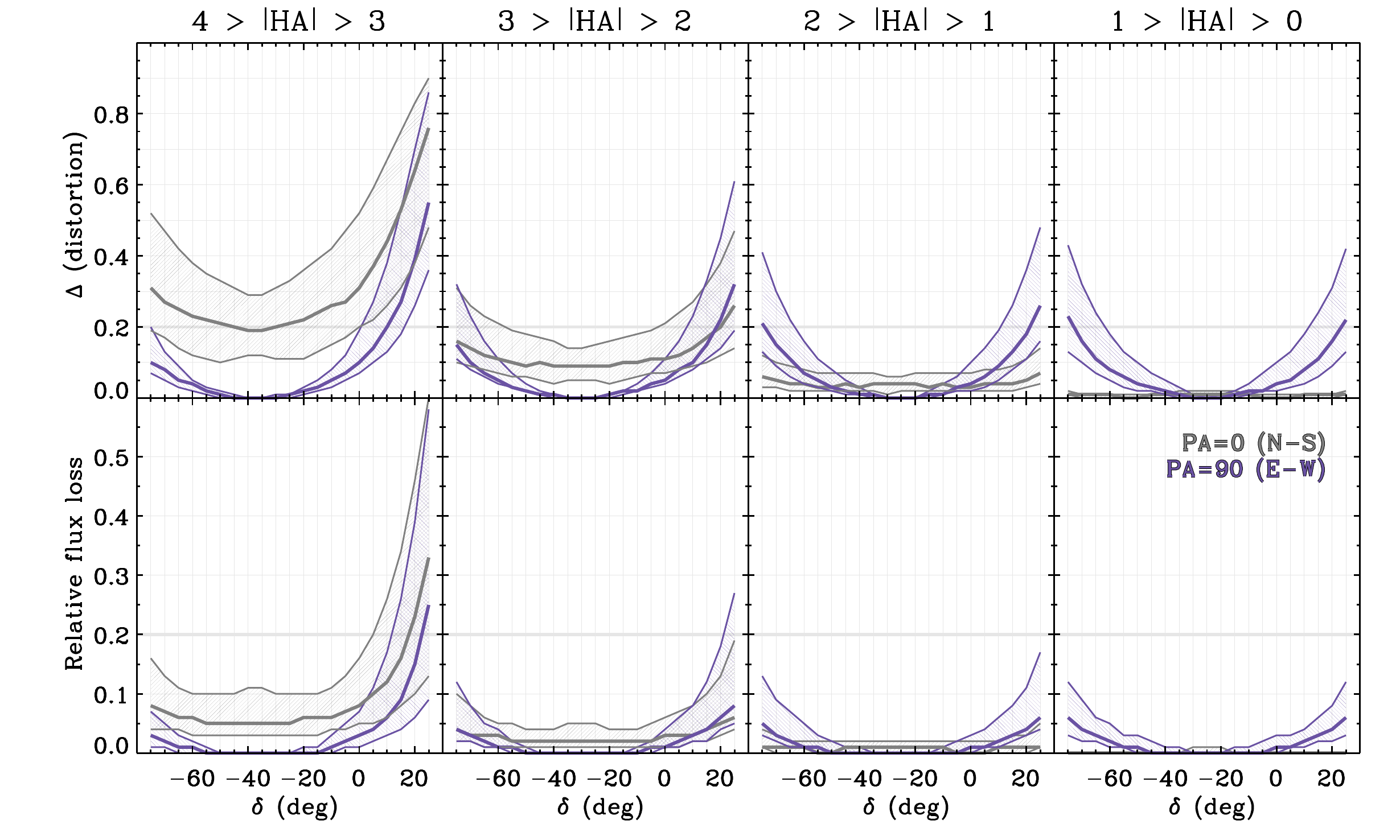}
\caption{Same as Fig.\,\ref{fig:hrblue}, but for the LR\_red grism.}
\label{fig:lrred}
\end{figure*}

Figure\,\ref{fig:slitloss_ex} shows an example of the output from our fiducial simulations\,\footnote{Source codes are available from the authors upon request.}. 
Each panel shows, for the nine different slit positions across the VIMOS FOV, the output spectra obtained after a one-hour-long integration (-3 $<$ HA (h) $<$ -2) on a $\delta$ = 0 deg field, and using the MR grism. 
The slit positions are fixed with respect to the centre of the FOV. At the meridian, $\Delta$x and $\Delta$y correspond to $\Delta\alpha$ and $\Delta\delta$, respectively.
Solid (dash-dotted) lines correspond to slits oriented along the N-S (E-W) direction at meridian crossing. The dotted lines indicate the fiducial maximum flux discussed in Sect.\,\ref{sect:model}. 
For each slit position we also provide the corresponding values for the two figures-of-merit with which we characterise the results of the simulations (see Table\,\ref{table:model}). 
The first one is the relative flux loss:
\begin{equation}\label{eq:flux}
f = 1 -  [\int_{\lambda_{min}}^{\lambda_{max}} \! f_{o}(\lambda)\,d\lambda~/ \int_{\lambda_{min}}^{\lambda_{max}} \! f_{i}\,d\lambda],
\end{equation}
\noindent
where $f_{o}(\lambda)$ is the output spectrum, and $\int_{\lambda_{min}}^{\lambda_{max}} \! f_{i}\,d\lambda$ is the fiducial maximum flux from the (flat) input spectrum. $f$ simply corresponds to the percentage flux reduction within the $[\lambda_{min},\lambda_{max}]$ wavelength range of a given grism (see Table\,\ref{table:vimos_setups}).
The second parameter we investigate is the spectral distortion:
\begin{equation}\label{eq:distortion}
\Delta = 1- f_{o,min}/f_{o,max}, 
\end{equation}
\noindent
where $f_{o,min}$ and $f_{o,max}$ are the minimum and maximum fluxes of the output spectrum, respectively.
Figure\,\ref{fig:slitloss_ex} shows that $\Delta$ provides supplementary information. Even though in this example the two slit orientations result in very similar median slit losses across the FOV ($f = 0.02$ and $f = 0.04$ for the E-W and N-S, respectively), the E-W alignment provides more stable results across the FOV, and a much lower median spectral distortion ($\Delta = 0.13$ vs. $\Delta = 0.22$). This should therefore be the preferred orientation in this case.

Figures\,\ref{fig:hrblue}-\ref{fig:lrred}  illustrate the final results for the entire set of simulations for each of the six grisms. In all panels the solid curves show the minimum, median and maximum  flux losses (lower row) and spectral distortions (upper row) for the nine simulated slits as a function of target declination. The shaded regions encompass all the output values. The plots show the effects for the two different slit orientations at the meridian (N-S in grey and E-W in purple).
Each column corresponds to a one-hour-long integration with target hour angle as indicated on top.
We note that the behaviour of the curves is similar for all grisms, but both losses and distortions are significantly smaller towards the red end of the visible spectrum.
The general trends for the two slit orientations can be summarised as follows.

For the N-S (PA = 0 deg) orientation we find that at fixed HA there is a very weak dependence on declination (except for $|$HA$|$ $>$ 2\,h and the bluest wavelengths). 
For any given grism, there is a strong dependence on HA, such that larger distortions and flux losses occur at larger HAs. 
Both the amount of losses/distortions, and the dependence on declination, increase for bluer wavelengths.
The minimum of the loss/distortion distributions increases and moves towards southern declinations  at larger HAs. 

On the other hand, for the E-W (PA = 90 deg) orientation, we see that at fixed HA there is a very strong dependence on declination, but the behaviour flattens towards redder wavelengths. 
For any given grism, there is very little dependence on HA (except for extreme declinations). 
Finally, the dependence on declination of losses/distortions increases towards bluer wavelengths.
The minimum of the loss/distortion distributions slightly decreases and moves towards southern declinations at larger HAs.

\begin{figure}\centering
\includegraphics[width=.5\textwidth]{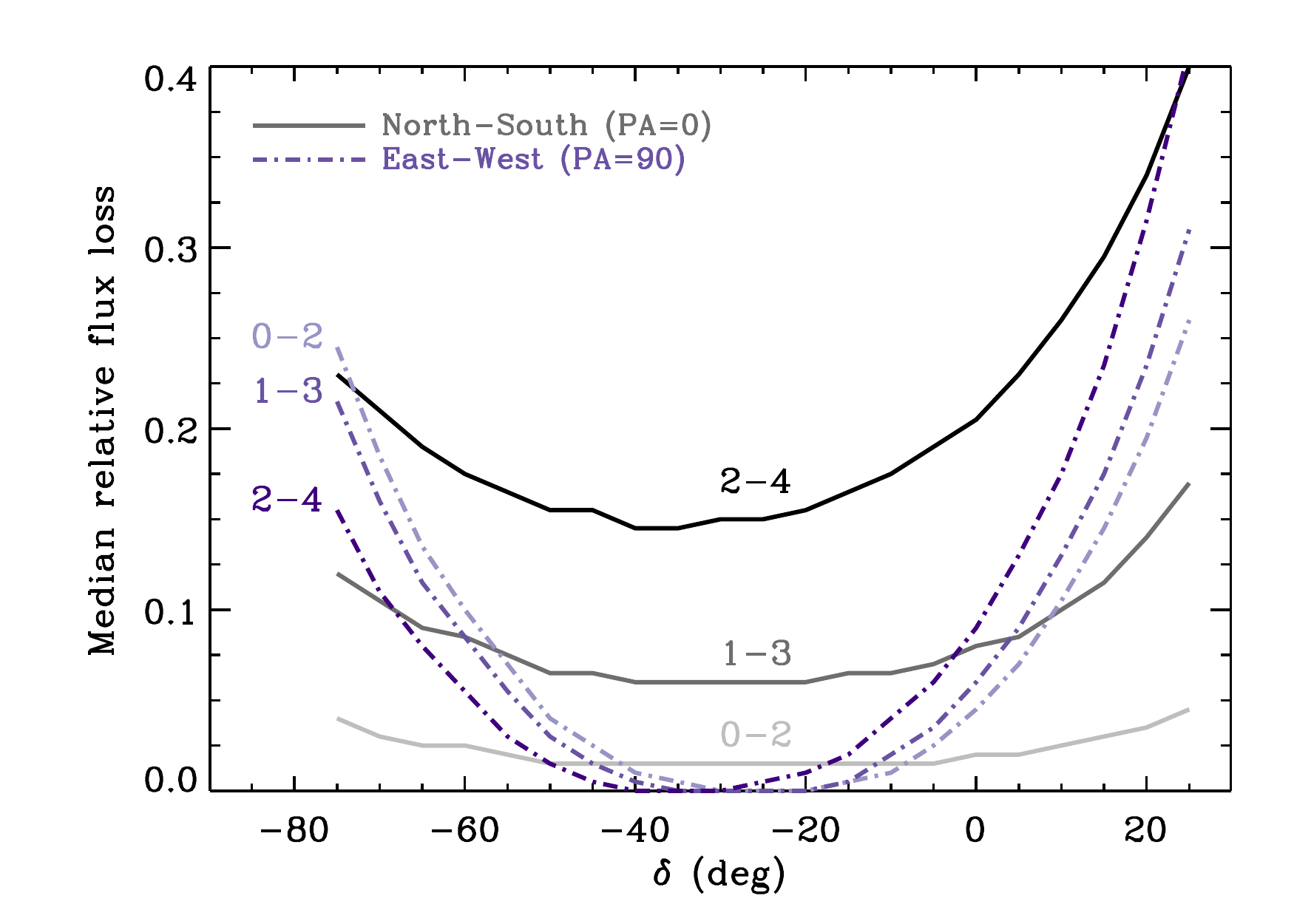}
\caption{Median percentage flux loss as a function of declination for two-hour-long integrations using the LR-blue grism, and spanning a range of hour angles (0-2 h, 1-3 h, and 2-4 h). Solid (dot-dashed) lines show the results for slits that are oriented along (perpendicular to) the dispersion direction at the meridian.}
\label{fig:manyhours}
\end{figure}

All these results correspond to the fiducial one-hour long integrations, which is the typical on-target exposure time of service-mode MOS OBs in VIMOS. In Fig.\,\ref{fig:manyhours} we show that these trends also hold for longer integrations. Curves show, for the LR-blue grism,  the median relative flux loss as a function of declination for two-hour-long integrations spanning a variety of hour angles -- namely 0-2\,h, 1-3\,h, and 2-4\,h. Solid (dot-dashed) lines show the results for slits that are oriented along (perpendicular to) the dispersion direction at the meridian. 
Naturally, the curve for a four-hour-long exposure before/after the meridian would simply be the mean between the 0-2\,h and the 2-4\,h ones.
The fact that slit losses are strongly dependent on hour angle for the N-S orientation, but almost insensitive to HA (at a given target declination) when aligning the slits along the E-W direction, has important implications for the scheduling of observations. 

\subsection{Beyond the two-hour angle rule}

\begin{figure*}\centering
\includegraphics[width=1.\textwidth]{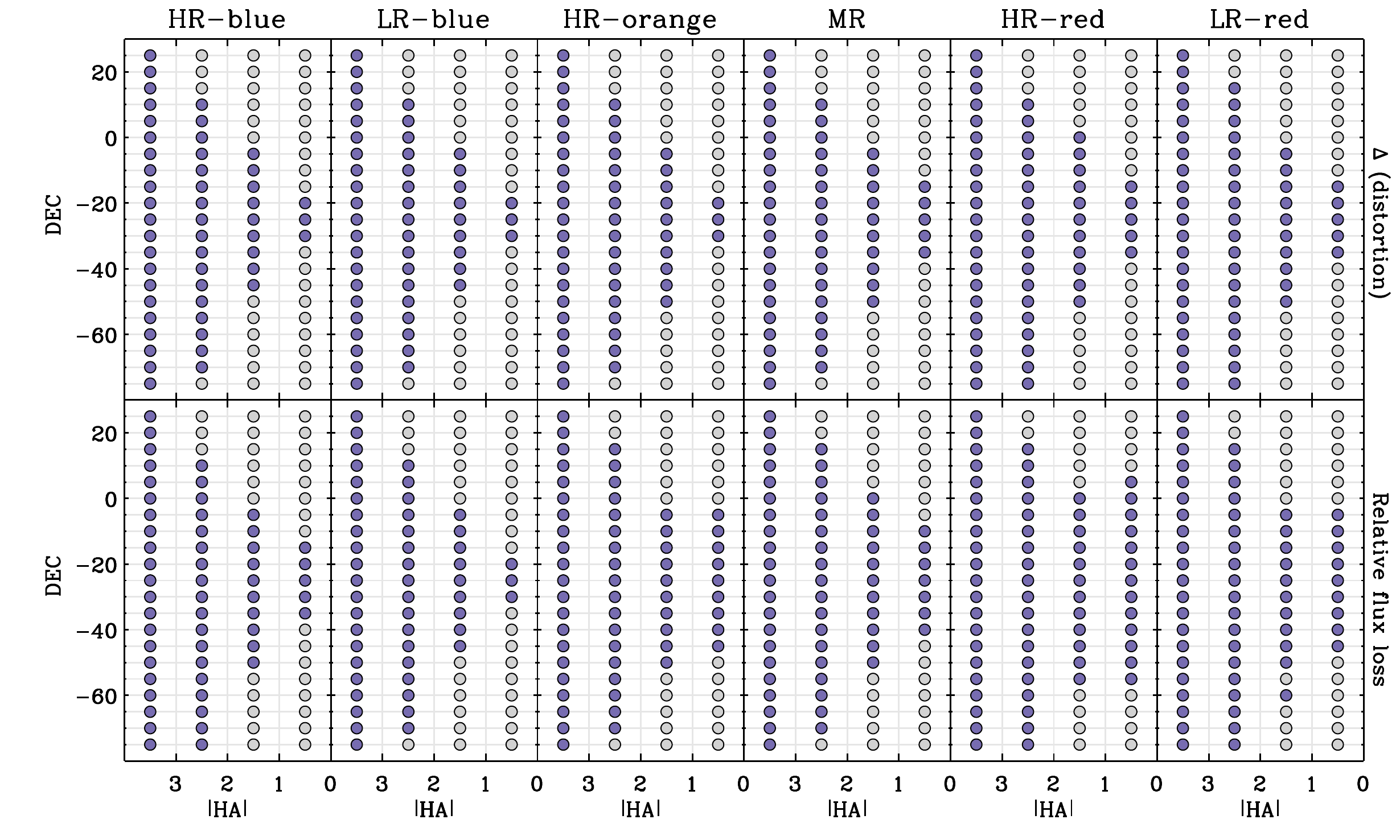}
\caption{Filled circles show the declination-hour angle pairs for which the median spectral distortion (top row) or median flux loss (bottom row) across the VIMOS FOV are lower with slits oriented along the N-S (grey symbols) or E-W (purple symbols) directions at meridian.}
\label{fig:nsew}
\end{figure*}

\begin{figure*}\centering
\includegraphics[width=1.\textwidth]{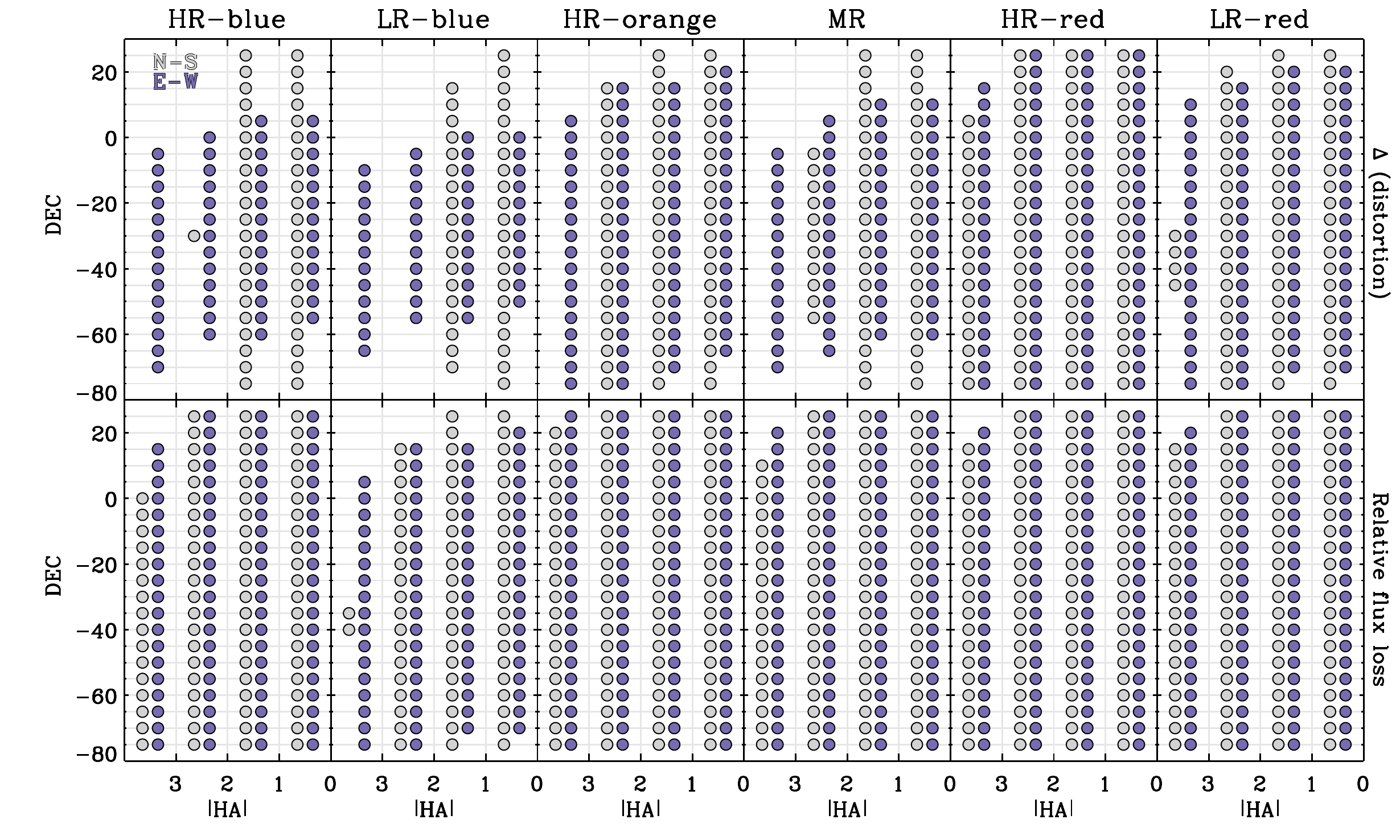}
\caption{Filled circles show the declination-hour angle pairs (colour-coded according to slit orientation) for which the median spectral distortion (top row) or median flux loss (bottom row) remain below the 20 per cent value during a one-hour-long integration.}
\label{fig:summary}
\end{figure*}

The trends previously discussed are well illustrated in Figure\,\ref{fig:nsew}. Here we colour-code the declination-hour angle pairs according to whether the median spectral distortion, or median flux loss, are lower with slits oriented along the N-S (grey symbols) or E-W (purple symbols) directions at meridian crossing. Each column summarises the results for all the VIMOS grisms, always assuming one-hour-long integrations.  
The tendencies previously discussed are perhaps more evident in this Figure, where it is shown that the E-W orientation is progressively preferred at larger hour angles. It is the favoured orientation for all declinations at HA $\gtrsim$ 3\,h, but only optimal for fields culminating at small zenith distances at HA $\lesssim$ 1\,h.
However, even if one of the orientations is formally superior to the other one,   slit losses and distortions can be unacceptably large even for the preferred orientation.
Additionally, a detailed inspection of Figs.\,\ref{fig:hrblue}-\ref{fig:lrred} indicates that in several cases the differences between orientations are almost negligible -- for all practical purposes they both perform equally well in these cases.
Following the previous work by Cuby et al. (1998), we set the tolerance level for losses/distortions at 20 per cent. In Fig.\,\ref{fig:summary} we similarly show the declination-hour angle pairs (again colour-coded according to slit orientation) for which the median spectral distortion, or median flux loss, across the VIMOS FOV remain below this tolerance value during a one-hour-long integration.


From these two Figures it is evident that the optimal slit alignment for fields culminating at small (large) zenith distances is the one that follows the E-W (N-S) direction at meridian crossing. The N-S orientation suffers from severe slit losses, and especially distortions, outside the two-hour interval from the meridian. The effect is particularly strong for the blue grisms, and turns milder towards redder wavelengths.
The E-W orientation, on the other hand, allows observations to be extended up to much larger hour angles -- up to four hours for certain fields, according to our fiducial model. 
However, as discussed in Sect.\,\ref{sect:deviations}, these are certainly lower limits to the actual losses and distortions.
We therefore prefer to be conservative, and recommend to limit service-mode observations with VIMOS to within three hours from the meridian even when using slits oriented E-W.


%
%

\section{ A note on the impact of field differential refraction on widefield multiobject spectrographs}\label{sect:fdr}

As discussed during this work, slit losses are the result of the combined effect of atmospheric dispersion and field differential refraction.
The former effects is usually dealt with ADCs, which are a common feature in most upcoming multiobject spectrographs. 
The latter is however an achromatic effect, and as a result dominates slit losses at the red end of the visual spectrum. Naturally, it can not be corrected with ADCs, and is therefore always present in widefield MOS observations. 

\begin{figure}
\includegraphics[width=.5\textwidth]{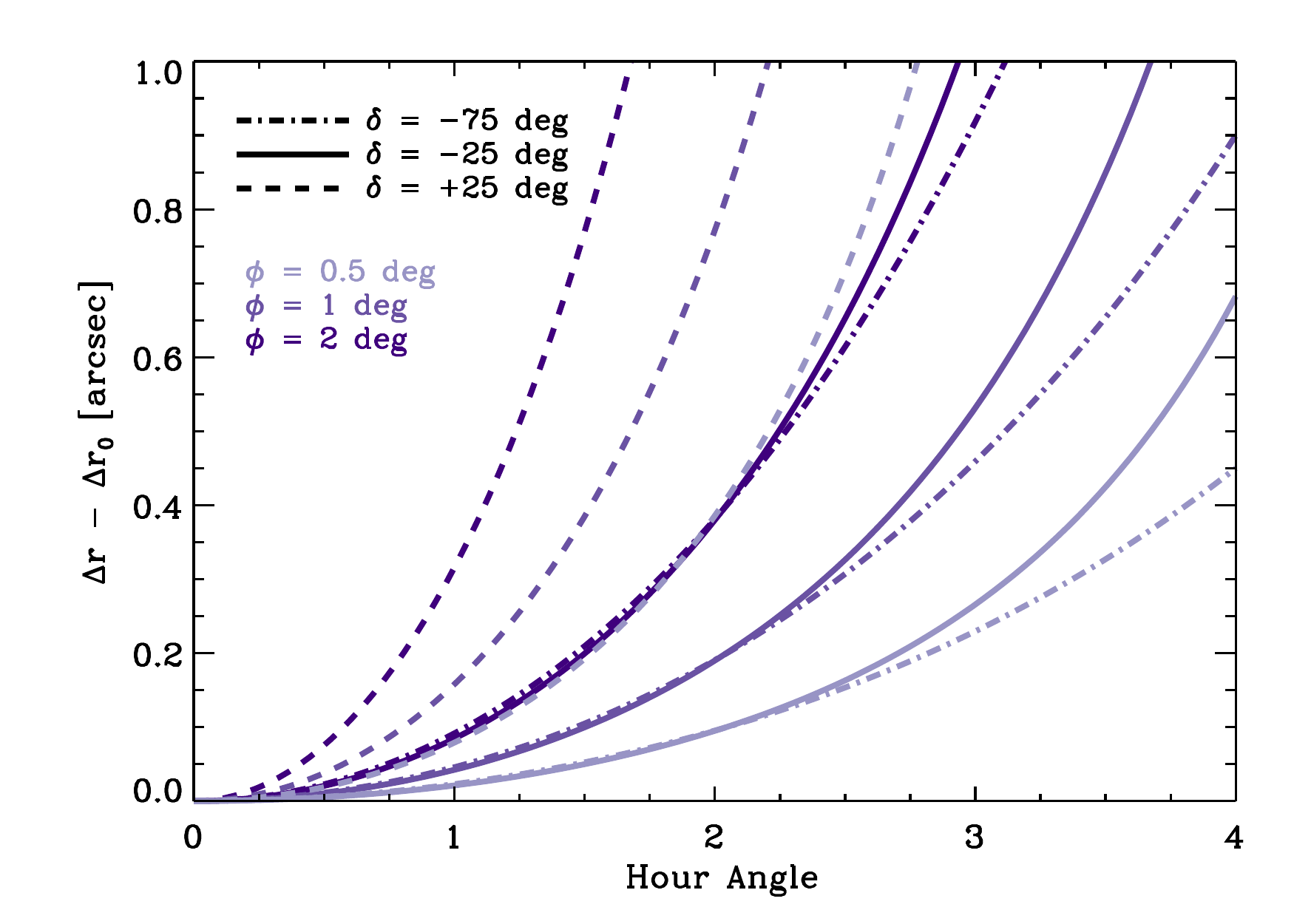}
\caption{Curves show the incremental field differential refraction for three different instrument FOVs ($\phi = 0.5, 1, 2$ deg), and three different fields ($\delta = -75, -25, + 25$ deg) observed from Paranal. All deviations are shown relative to the effect at HA = 0\,h. The effect starts to become significant ($\gtrsim$ 0.1 arcsec) even for HA $\sim$ 1\,h for fields culminating to the North.}
\label{fig:fdr}
\end{figure}

The field differential refraction across a given field can be approximated by 

\begin{equation}\label{eq:fdr}
\Delta r \simeq (n-1)\,\phi\,X^{2}, 
\end{equation}

\noindent
where $X$ is again the airmass, $\phi$ is the diameter of the FOV, and $n$ is the index of refraction of air, of the order of $(n-1) \approx 2\times10^{-4}$ in the optical at the typical elevation of astronomical observatories. 
Equation\,\ref{eq:fdr} shows that, for FOVs similar to that of VIMOS, $\Delta r \simeq 0.25$ arcsec even at moderate airmasses -- and the effect increases linearly with $\phi$. 
This makes field differential refraction a non-negligible effect for most widefield multiobject spectrographs that can only be compensated by the scheduling of the observations. 
In the case of multislit spectrographs like VIMOS, the only alternative is to limit the observations to the airmass range where differential refraction is not critical.
In fibre-fed spectrographs the fibres can be configured at the beginning of an observation, so that the mean effect during the entire exposure is minimised. The fibres will eventually need to be reconfigured to avoid a significant flux reduction, and the amount of time during which any given field can be observed without significant losses depends strongly on its declination.
If we substitute for the airmass term in Eq.\,\ref{eq:fdr}, and  write the field differential refraction as a function of the hour angle for a given field:

\begin{equation}\label{eq:fdrha}
\Delta r \simeq (n-1)\,\phi\,[\alpha + \beta\,\mbox{cos}\,\mbox{HA}]^{-2},
\end{equation}

\noindent
where $\alpha = \mbox{sin}\,\varphi\,\mbox{sin}\,\delta$, and $\beta = \mbox{cos}\,\varphi\,\mbox{cos}\,\delta$.
In Fig.\,\ref{fig:fdr} we plot the \emph{incremental} deviation (i.e., with respect to the effect at HA = 0\,h) caused by differential refraction for three different instrument FOVs (diameters $\phi = 0.5, 1, 2$ deg), and three different fields ($\delta = -75, -25, + 25$ deg) observed from Paranal.
Naturally, the effect is more important at larger airmasses, and for instruments with large FOVs.

\begin{figure}
\includegraphics[width=.5\textwidth]{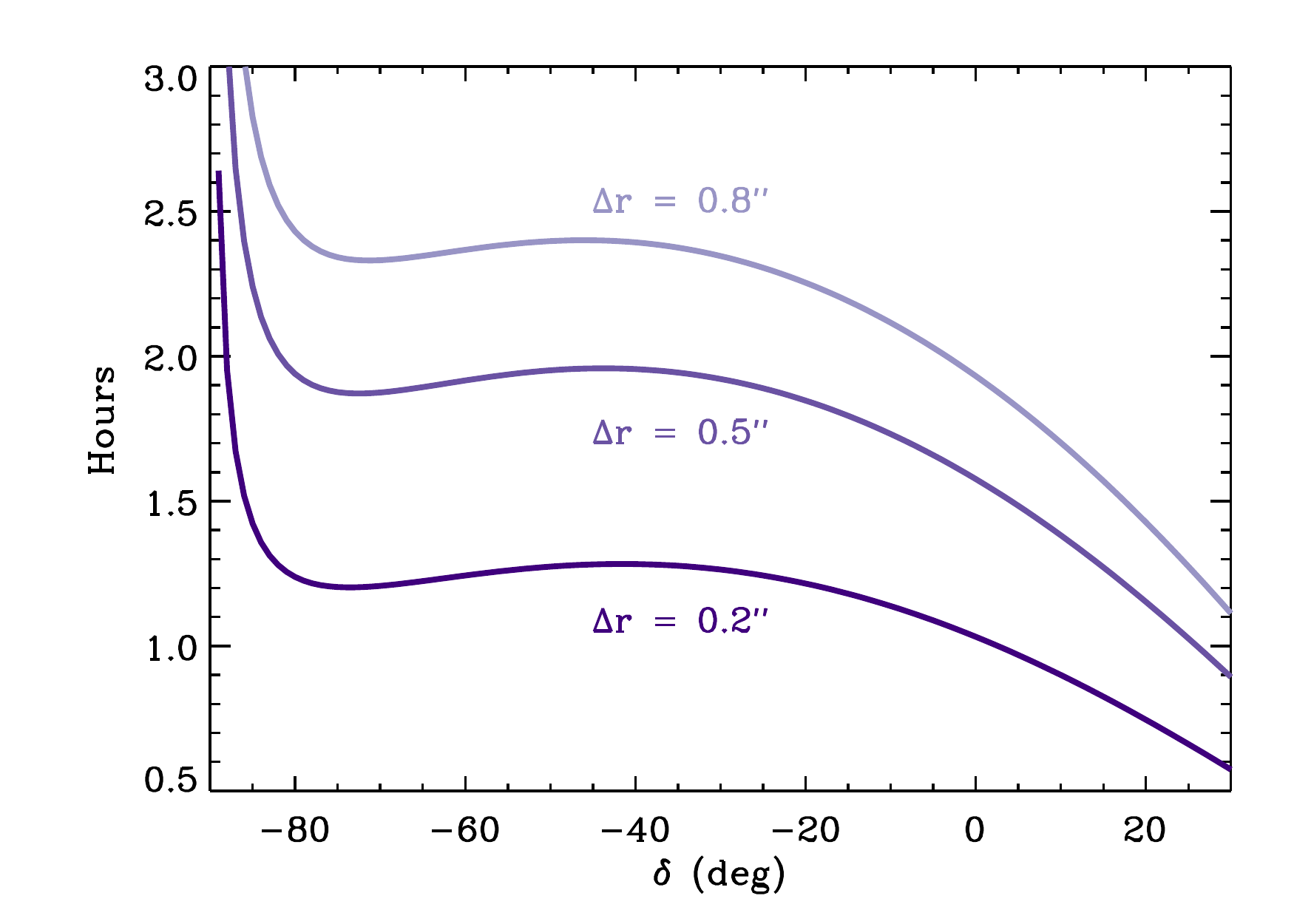}
\caption{Time needed for incremental field differential refraction to exceed a given deviation, as a function of target declination. These curves are computed for an instrument similar to 4MOST, a $\phi = 3$ deg FOV spectrograph at the latitude of Paranal. Southern fields can be observed for times longer than one hour without resulting in significant deviations. That is not the case for northern fields,  where  fibre reconfiguration will be needed within the first hour of integration.}
\label{fig:fdr_4most}
\end{figure}

Indeed, in Fig.\,\ref{fig:fdr_4most} we show, for a multiobject spectrograph featuring a 3 deg diameter FOV (e.g., 4MOST, \citealt{deJong2011}), the amount of time needed for incremental field differential refraction to exceed a given deviation, as a function of target declination (assuming Paranal's latitude).
In general, southern fields can be observed for times longer than one hour without resulting in significant deviations -- hence with minimal flux reduction. On the other hand, the deviation rapidly exceeds 0.2 arcsec for northern fields, and fibre reconfiguration will be needed before the first hour of integration.
All-sky type MOS surveys must therefore carefully plan the scheduling of observations in order to minimise these effects. 
 
%
%

\section{Discussion and conclusions}\label{sect:conclusions}

We have shown that the optimal slit alignment in widefield MOS observations is a complex function of target declination and hour angle (Fig.\,\ref{fig:nsew}).
The traditional orientation of slits along the parallactic angle at the meridian provides very stable results for observations within two hours from the meridian. A setup where slits are oriented \emph{perpendicular} to the parallactic angle at the meridian is however preferred in the case of fields culminating at low zenith distances, or generally when observations extend to $|$HA$| >$ 2\,h.
This result can readily be understood from the following qualitative line of arguments.
As discussed by \citet{Szokoly2005}, slit losses depend on the interplay between two effects. 
The first one is independent of the slit orientation, as both atmospheric dispersion and differential refraction are larger at increasing airmasses. 
On the other hand, the angle between the slit and the parallactic angle is constantly changing.
When slits are oriented along the parallactic angle at the meridian (N-S), the angle between the slits and the parallactic angle tends to increase, and the two effects amplify each other.
Conversely, when slits are oriented perpendicular to the parallactic angle at the meridian (E-W) --and dispersion is always minimal--, the slits get systematically closer to the parallactic angle at higher airmasses, so that a large fraction of the dispersed light falls back into the slits.
This is what creates the opposing trends with declination and hour angle depicted in Figs.\,\ref{fig:hrblue}-\ref{fig:lrred}: the N-S orientation depends strongly on HA, but not so much on $\delta$, while the strong dependence of  the E-W orientation on $\delta$ is almost invariant to changes on HA. 

\begin{figure}
\includegraphics[width=.5\textwidth]{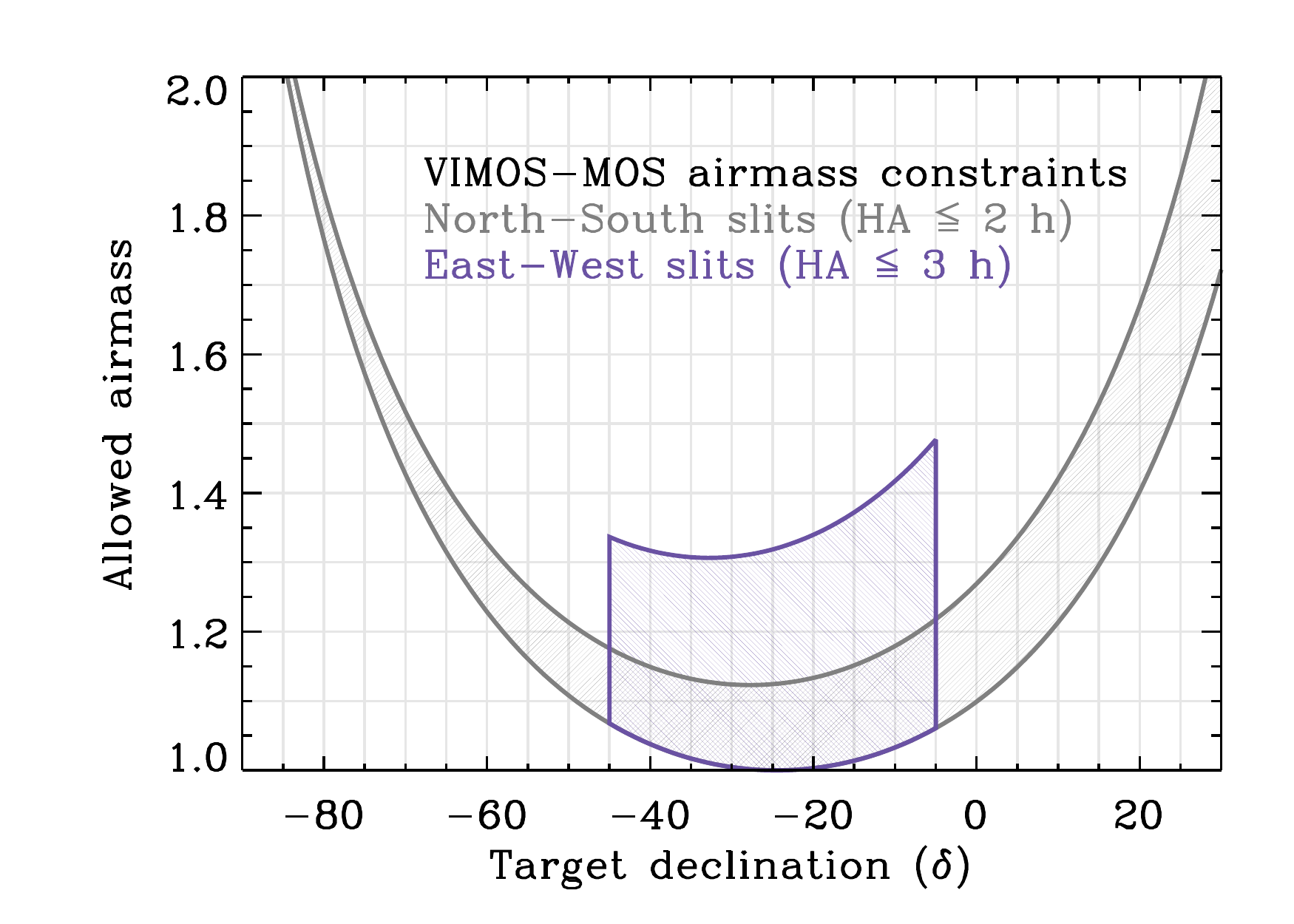}
\caption{VIMOS airmass constraints for MOS observing blocks. The two different shaded areas correspond to the limits for fields that can be observed with slits having N-S orientations at meridian (grey), or E-W orientations (purple). The lower curve in each case corresponds to the physical limit for targets of that declination.}
\label{fig:airmass}
\end{figure}

This has important implications for the scheduling of observations. 
Night-long observations of a single field (visitor mode) will generally benefit from using the E-W orientation for slits at the meridian.
All-sky or service mode observations, however, require a more elaborate planning that depends on the target and the hour angle of the observations.
This is a particularly relevant aspect for the forthcoming spectroscopic public surveys with VIMOS, where observations will be carried out in visitor mode.
We conclude that the two-hour angle rule, together with the default N-S slit orientation, provide the most stable results for service observations with VIMOS, with slit losses and spectral distortions below 20 per cent -- and almost independent of target declination. This should always be the preferred option for users having targets at $\delta \gtrsim -5$ deg or $\delta \lesssim -45$ deg. 
However, for targets within the $-45 \lesssim \delta \lesssim -5$ deg range, the E-W orientation results in comparable or even reduced slit losses. This slit orientation allows for observations to go past the two-hour angle rule, and  be effectively extended up to $|$HA$| = 3$ hours, hence making it a preferred option. 
This holds for all grisms currently offered in VIMOS, provided the acquisition is done with a filter that closely matches the grism wavelength range.

Figure\,\ref{fig:airmass} shows the new airmass constraint limits for MOS OBs. They have been significantly relaxed for fields culminating at small zenith distances, thus increasing target observability. This shall enhance the efficiency of operations, and speed up the completion of programmes.
Even though the results presented here are specifically tailored to VIMOS, these general recommendations apply to all current and future widefield multislit spectrographs (e.g., Magellan/IMACS, \citealt{dressler2011}; GMT/GMACS, \citealt{depoy2012}).
They are already in place since September 2013 for VIMOS-MOS observations \citep{rsj2013}.

%
%

\begin{acknowledgements}
The authors thank the referee, Katherine Roth, for a very detailed review of the manuscript that has significantly improved the quality of the paper.
We acknowledge Amelia Bayo, Caroline Foster, Claudio Melo, and Alain Smette for interesting discussions and feedback. Finally, we thank Julio Navarrete for providing us with the latest measurements of atmospheric pressure and temperature at Paranal.
\end{acknowledgements}

\bibliographystyle{aa}
\bibliography{../../rsj_references.bib}

%
%



%
%

\end{document}